\documentclass[twocolumn,showpacs,preprintnumbers,amssymb,pra]{revtex4}
\usepackage{graphicx}
\usepackage{dcolumn}
\usepackage{color}
\usepackage{bm}
\begin{document}

\title {Quantum teleportation using non-orthogonal entangled channels}
\author{Satyabrata Adhikari}
\altaffiliation{satyabrata@bose.res.in}
\affiliation{S. N. Bose National Centre for Basic Sciences,
Salt Lake, Kolkata 700 098, India}
\author{A. S. Majumdar}
\altaffiliation{archan@bose.res.in}
\affiliation{S. N. Bose National Centre for Basic Sciences,
Salt Lake, Kolkata 700 098, India}
\author{Dipankar Home}
\altaffiliation{dhome@bosemain.boseinst.ac.in}
\affiliation{CAPSS, Dept. of Physics, Bose Institute, Salt Lake,
Kolkata-700091, India}
\author{A. K. Pan}
\altaffiliation{apan@bosemain.boseinst.ac.in}
\affiliation{CAPSS, Dept. of Physics, Bose Institute, Salt Lake,
Kolkata-700091, India}
\author{P. Joshi}
\altaffiliation{writepank@gmail.com}
\affiliation{Department of Physics, Pune University, Pune 411007, India}

\date{\today}

\vskip 0.5cm
\begin{abstract}
We study quantum teleportation with the resource of non-orthogonal qubit
states. We first extend the standard teleportation protocol to the case
of such states. We investigate how the loss of teleportation fidelity
resulting for the use of non-orthogonal states compares to a similar
loss of fidelity when noisy or non-maximally entangled states as used
as teleportation resource. Our analysis leads to certain interesting
results on the teleportation efficiency of both pure and mixed non-orthgonal
states compared to that of non-maximally entangled and mixed states.
\end{abstract}


\maketitle

\section{Introduction}

Quantum teleportation is an important and vital quantum
information processing task where an arbitrary unknown quantum
state can be replicated at a distant location using previously
shared entanglement and classical communication between the sender
and the receiver. A remarkable application of entangled states
having many ramifications in information technology, quantum
teleportation can also be combined with other operations to
construct advanced quantum circuits useful for information
processing \cite{nc}. The original idea of teleportation
introduced by Bennett et al. \cite{bbcjpw} is implemented through
a channel involving a pair of particles in a Bell State shared by
the sender and the receiver and at the end of the protocol an
unknown input state is reconstructed with perfect fidelity at
another location while destroying the original copy. 

For implementing  teleportation one ideally needs maximally entangled
two-qubit states, i.e., singlet states. But in a typical
experiment it is very difficult to prepare singlet states because
the preparation is never perfect. As a result, one may generally
have to deal with non-maximally entangled states or mixed states. 
So it becomes
necessary to generalize the idea of using the maximally entangled
states as quantum channels to the case of non-maximally entangled
or noisy channels between two distant partners \cite{bbps}. To
switch over from a maximally entangled state to a non-maximally
entangled state or a mixed state, one has to pay a 
price in terms of the loss of
teleportation fidelity. For the case of mixed states, there are several
works in the literature on schemes of teleportation using different 
categories of mixed states (see, for example \cite{horodecki1,kim})
for Werner states \cite{werner}, and \cite{ishizaka} for a broader class
of mixed states \cite{munro}). For non-maximally entangled channels the 
loss of fidelity of teleportation could be compensated by
schemes for probabilistic teleportation \cite{ap,ap1}.
Probabilistic teleportation
schemes have  been developed to teleport N qubits using directly N 
non-maximally entangled channels \cite{gr}.

In quantum information theory, the entanglement of
orthogonal states has received much attention. However,
non-orthogonal states could also act as potentially 
useful resources for information processing. 
Examples of entangled non-orthogonal states are readily available, {\it viz.}
entangled coherent states fall under this category. A
Schr$\ddot{o}$dinger cat state which has been proposed using
$\textit{SU(2)}$ coherent states forms a particular realization of the
entanglement of non-orthogonal $\textit{SU(2)}$ coherent states
\cite{ys}. The theory of non-orthogonal quantum states has developed with
the study of Schmidt decomposition for a two particle system
involving non-orthogonal states \cite{msm}. The utility of
non-orthogonal quantum states for
cryptographic purposes \cite{fuchs1} is appreciated because nonorthogonal 
quantum states cannot
be distinguished with perfect reliability \cite{fuchs}. Any
attempt to do so (even imperfectly) imparts a disturbance to them
\cite{fp}. Studies on non-orthogonal quantum
states which form a overcomplete basis in the context of
teleportation have been undertaken and a realistic protocol
for the continuous variable teleportation of a coherent state has been
suggested \cite{bk}.  The
continuous-variable teleportation protocol was implemented in an
experiment that teleported a coherent state of an
optical-frequency electromagnetic mode with fidelity $0.58\pm
0.02$ \cite{fsbfkp}. Two other experiments have improved the
experimental fidelity of the teleported coherent state to values
of $0.64\pm 0.02$ \cite{btbsrbsl} and $0.61 \pm 0.02$ \cite{zgclk}.
This protocol has been developed further recently \cite{yf}.

Given that teleportation protocols have been proposed with non-maximally 
entangled, mixed, as well as non-orthogonal entangled channels, a systematic
study is in order to evaluate their comparative performance in terms of
their respective teleportation fidelities. With such a  motivation
in this work, we first reformulate the original quantum teleportation
protocol \cite{bbcjpw} in terms of  non-orthogonal quantum states in a
two-dimensional Hilbert space. Since the introduction of an infinitesimal amount
of non-orthogonality decreases the amount of entanglement in a
bipartite system, it would be interesting to see how the
introduction of non-orthogonality in the system affects the
fidelity of teleportation. The issue as
to whether the average teleportation fidelity can be made to
increase when a non-orthogonal two qubit entangled system is used
compared to the case when either a non-maximally orthogonal entangled state,
or a mixed state
is used as a teleportation channel, could be important for practical
purposes. Further, since even among mixed states there exist two
distinct categories, {\it viz.} maximally entangled \cite{werner,munro} or
non-maximally entangled \cite{ishizaka,qic} states, a similar 
comparison could be performed for a mixed entangled non-orthogonal
channel with respect to a non-maximally entangled mixed state which
has been observed to yield a better teleportation fidelity than some
other mixed states \cite{qic}.

The plan of this paper is as follows.
In the next section we present our protocol for teleportation using
a non-orthogonal qubit state as the channel. In Section III we compare 
the performance (fidelity of teleportation) of this non-orthogonal
channel with that of a non-maximally entangled channel, and also the Werner
state used as a teleportation channel. We next perform
a similar study for the case of mixed states in Section IV. Here we
compute the fidelity of teleportation for a non-orthogonal mixed state
and compare it with that of a non-maximally entangled mixed state.  
The summary of our results is presented in Section V.

\section{Non-orthogonal entangled state and teleportation}

A bipartite entangled state can, in general, be
written as
\begin{eqnarray}
|\alpha\rangle^{AB}=\mu |\alpha\rangle^{A}|\beta\rangle^{B}+\nu
|\gamma\rangle^{A}|\delta\rangle^{B} \label{gen}
\end{eqnarray}
where the state vectors $|\alpha\rangle^{A}$ and
$|\gamma\rangle^{A}$ for system A and $|\beta\rangle^{B}$ and
$|\delta\rangle^{B}$ for system B represents the linearly
independent non-orthogonal states that span a two-dimensional
subspace of each Hilbert spaces. The parameters $\mu$ and $\nu$
are the complex coefficients. A bipartite entangled state
involving non-orthogonal states would have the property that the
overlaps $\langle \alpha|\gamma\rangle^{A}$ and $\langle
\delta|\beta\rangle^{B}$ are non-zero.

Let us consider two-dimensional subspace of the Hilbert spaces
$H_{A}$ and $H_{B}$ spanned by the linearly independent
non-orthogonal quantum states $|\alpha\rangle$ and
$|\beta\rangle$.
 The quantum states $|\alpha\rangle$ and
$|\beta\rangle$ are non-orthogonal in the sense that their inner
product is non-vanishing i.e. $\langle\alpha|\beta\rangle\neq0$.
Since, in general, $\langle\alpha|\beta\rangle$ is complex so we
can assume $\langle\alpha|\beta\rangle=re^{i\theta}$, where the
real parameters $r$ and $\theta$ respectively denotes the modulus
and argument of the complex number.
Let us choose the normalized non-orthogonal basis vectors
$|\alpha\rangle$ and $|\beta\rangle$ as
\begin{eqnarray}
|\alpha\rangle=\left(\begin{matrix}{0 \cr
1}\end{matrix}\right),~~~~ |\beta\rangle=\left(\begin{matrix}{
re^{i\theta} \cr N_{\beta}}\end{matrix}\right) \label{nob}
\end{eqnarray}
where $N_{\beta}=\sqrt{1-r^{2}}$.
Using Gram-Schmidt ortho-normalization procedure, we transform
non-orthogonal basis vectors $|\alpha\rangle$ and $|\beta\rangle$
into normalized orthogonal basis vectors $|0\rangle$ and
$|1\rangle$ as
\begin{eqnarray}
|0\rangle=|\alpha\rangle,~~~~~|1\rangle=N_{G}[|\beta\rangle-\langle\alpha|\beta\rangle|\alpha\rangle]
\label{ob}
\end{eqnarray}
where $N_{G}=\frac{1}{N_{\beta}}=\frac{1}{\sqrt{1-r^{2}}}$.
Here we note that the state $|1\rangle$ contains implicitly the
information of non-orthogonality of the original non-orthogonal
system. So we may now proceed with usual orthogonal
basis vectors $|0\rangle$ and $|1\rangle$.

A bipartite entangled state in the non-orthogonal basis can be written
as
\begin{eqnarray}
|\Psi\rangle_{ab}=
N_{1}[|\alpha\rangle|\beta\rangle+|\beta\rangle|\alpha\rangle]
\label{noes}
\end{eqnarray}
where $N_{1}=\frac{1}{\sqrt{2(1+r^{2})}}$ is the normalization
constant.
In terms of the orthogonal basis vectors $|0\rangle$ and $|1\rangle$,
Eq.(\ref{noes}) can be re-expressed as
\begin{eqnarray}
|\Psi\rangle_{ab}= \frac{N_{1}}{N_{G}}(|0\rangle|1\rangle+
|1\rangle|0\rangle)+2N_{1}\langle\alpha|\beta\rangle|0\rangle|0\rangle
\label{oes}
\end{eqnarray}
The amount of entanglement contained in the bipartite entangled
state $|\Psi\rangle_{ab}$ can be quantized by the 
concurrence $(C)$  given by
\begin{eqnarray}
C(|\Psi\rangle_{ab})=2\sqrt{det(\rho_{a})} \label{conc.}
\end{eqnarray}
where
\begin{eqnarray}
\rho_{a}=Tr_{2}(|\Psi\rangle_{ab})=\left(\begin{matrix}{\frac{1+3r^{2}}{2(1+r^{2})}
& \frac{re^{i\theta}\sqrt{1-r^{2}}}{1+r^{2}}\cr
\frac{re^{-i\theta}\sqrt{1-r^{2}}}{1+r^{2}}
&\frac{1-r^{2}}{2(1+r^{2})}}\end{matrix}\right)\label{red.den}
\end{eqnarray}
From Eq.(\ref{conc.}) and (\ref{red.den}) we have
\begin{eqnarray}
C(|\Psi\rangle_{ab})= \frac{1-r^{2}}{1+r^{2}},~~~~0\leq r\leq 1
\label{finalconc.}
\end{eqnarray}
The parameter $r$ is a measure of the
non-orthogonality. From the expression (\ref{finalconc.}) we find that
$C$ is a decreasing function of $r$ and hence as the amount of
non-orthogonality increases, the amount of entanglement in a
bipartite system decreases and it goes to zero when the non-orthogonal
parameter $r$ tends to 1. The maximum amount of entanglement is
achieved when $r=0$, i.e., when the basis state vectors are
orthogonal to each other.

Let us now formulate the teleportation protocol between two parties
Alice and Bob,
with the input state prepared by a third party Charlie. He 
sends the prepared state to Alice. In this transmission we assume
that there is no distortion of the input state. Since the input
state is given by the third party, Alice have no knowledge about the
received state which she wants to teleport, and this arbitrary
state is given by
\begin{eqnarray}
|\phi\rangle_{1}= x |0\rangle + y |1\rangle \label{input}
\end{eqnarray}
where $x^{2}+y^{2}=1$.
Let us assume that two distant partners Alice and Bob  share a two
qubit entangled state $|\Psi\rangle_{ab}$ given by Eq.(\ref{oes}).
The particles $a$ and $b$ are with Alice and Bob, respectively.
The state $|\Psi\rangle_{ab}$ acts as a quantum
teleportation channel whose entanglement depends on how much
non-orthogonality one introduces in the system.

Next, we combine the single qubit $\textit{1}$ and the two qubits
$\textit{a}$ and $\textit{b}$ using the tensor product and then express the
resulting three qubit system as a tensor product of a single qubit
$\textit{b}$ and the Bell basis involving the two qubits $\textit{1}$ and
$\textit{a}$, as
\begin{eqnarray}
&&|\chi\rangle_{1ab}= |\phi\rangle_{1}\otimes |\Psi\rangle_{ab} \nonumber\\
&=&
\frac{1}{\sqrt{2}}[|\Phi^{+}\rangle_{1a}(P_{+}|0\rangle_{b}+A|1\rangle_{b})
+|\Phi^{-}\rangle_{1a}(P_{-}|0\rangle_{b}+A|1\rangle_{b}){}\nonumber\\
&+& \hskip -0.2cm |\Psi^{+}\rangle_{1a}(Q_{+}|0\rangle_{b}+B|1\rangle_{b})
+|\Psi^{-}\rangle_{1a}(Q_{-}|0\rangle_{b}-B|1\rangle_{b})]
\label{input}
\end{eqnarray}
where $A=\frac{xN_{1}}{N_{G}}$, $B=\frac{yN_{1}}{N_{G}}$,
$P_{\pm}=N_{1}(2x re^{i\theta}\pm\frac{y}{N_{G}})$,
$Q_{\pm}=N_{1}(\frac{x}{N_{G}}\pm 2yre^{i\theta})$,
$|\Phi^{\pm}\rangle=\frac{1}{\sqrt{2}}(|00\rangle\pm|11\rangle)$,
$|\Psi^{\pm}\rangle=\frac{1}{\sqrt{2}}(|01\rangle\pm|10\rangle)$.

Since the qubits $\textit{1}$ and $\textit{a}$ are with Alice,  she performs
a Bell state measurement on her qubits and then sends the measurement
result to Bob expending two classical bits.
According to the received measurement result $|\Phi^{+}\rangle$,
$|\Phi^{-}\rangle$, $|\Psi^{+}\rangle$ or $|\Psi^{-}\rangle$, Bob
performs a suitable unitary operation on his
qubit $b$ as follows:\\
(i) If the measurement result is $|\Phi^{+}\rangle$,  Bob
operates $\sigma_{x}$ on his qubit.\\
(ii) If the measurement result is $|\Phi^{-}\rangle$,  Bob
operates $\sigma_{y}$ on his qubit.\\
(iii) If the measurement result is $|\Psi^{+}\rangle$, 
Bob operates the Identity operator $I$ (does nothing) on his qubit.\\
(iv) If the measurement result is $|\Psi^{-}\rangle$, Bob operates
$\sigma_{z}$ on his qubit.\\
Thereafter Bob transmits this
state to the third party Charlie whose task would be to measure
the efficiency of the teleportation protocol.

The teleportation fidelity is defined as
\begin{eqnarray}
F^{tel}=\sum_{j=1}^{4}P_{j}|\langle\phi|\xi_{j}\rangle|^{2}
\label{fid.}
\end{eqnarray}
where $|\phi\rangle$ is the input state and
$|\xi_{j}\rangle~(j=1,2,3,4)$ are the normalized single qubit
output states after the unitary transformations and
$P_{j}=tr(M_{j}|\xi_{j}\rangle\langle\xi_{j}|)$ denotes the
corresponding probability of getting the normalized output state
$|\xi_{j}\rangle$, where
$M_{1}=|\Phi^{+}\rangle\langle\Phi^{+}|$,$M_{2}=|\Phi^{-}\rangle\langle\Phi^{-}|$,
$M_{3}=|\Psi^{+}\rangle\langle\Psi^{+}|$,$M_{4}=|\Psi^{-}\rangle\langle\Psi^{-}|$.
In this case, the teleportation fidelity is found out to be
\begin{eqnarray}
F^{tel}= \frac{1-r^2(1-2y^2)^2}{1+r^2} \label{fid.1}
\end{eqnarray}
From Eq.(\ref{fid.1}), one can observe the
following points:\\
(i) If $y\rightarrow0$ or $y\rightarrow1$,  the fidelity of
teleportation of a qubit (lying in the neighborhood of classical
bit) via the non-orthogonal entangled state $|\Psi\rangle_{ab}$ as a
teleportation channel is given by
\begin{eqnarray}
F^{tel} \rightarrow \frac{1-r^2}{1+r^2} = C(|\Psi\rangle_{ab})
\label{fid.2}
\end{eqnarray}
The fidelity of teleportation (\ref{fid.2}) exceeds the classical
fidelity $\frac{2}{3}$ when the  parameter $r$ satisfies the inequality
$0\leq r <\frac{1}{\sqrt{5}}$.\\
(ii) If $y=\frac{1}{\sqrt{2}}$, i.e., if the qubit is in an equal
superposition of two classical bits,  the fidelity of
teleportation is given by
\begin{eqnarray}
F^{tel}_{1} =\frac{1}{1+r^2} \label{fid.3}
\end{eqnarray}
In this case, the fidelity (\ref{fid.3}) overtakes classical
fidelity $\frac{2}{3}$ when $0 \leq r < \frac{1}{\sqrt{2}}$.

Since the teleportation fidelity (\ref{fid.1}) is input state
dependent, it would be better to calculate the average fidelity
over all input states. The average teleportation fidelity over all
input states is given by
\begin{eqnarray}
F^{tel}_{av}= \frac{3-r^2}{3(1+r^2)} \label{av.fid}
\end{eqnarray}
Here we have obtained the effect of non-orthogonality on the
average teleportation fidelity when a non-orthogonal entangled
quantum channel is used in the teleportation protocol. Let us now
compare the results of this section with the case when a
non-maximally entangled state is used as a teleportation channel.

\section{Comparison of teleportation fidelities of a Non-orthogonal channel
with mixed and Non-maximally entangled channels}

One of the best known, and perhaps also the simplest example of a mixed
qubit state is the Werner state \cite{werner} which is a convex combination
of a pure maximally entangled state and a maximally mixed state. Since
the entanglement of the Werner state cannot be increased by any unitary
transformation, it can be
regarded as a maximally entangled mixed state \cite{ishizaka}. The efficiency
of the Werner state as a teleportation channel has been studied in detail
\cite{kim}. Let us here briefly recapitulate some of the essential features
relevant to our present study. The Werner state can be expressed as
\begin{eqnarray}
\rho^W = (1-p)|\psi^-\rangle\langle \psi^-| + \frac{p}{4}I
\label{wernerstate}
\end{eqnarray}
where $|\psi^-\rangle$ is the singlet state and the parameter $p$ lies between
$0$ and $1$, with $p=0$ denoting a maximally entangled pure state, while $p=1$
denotes the maximally mixed state. When the Werner state is used as a 
teleportation channel the average teleportation fidelity is given in
terms of the parameter $p$ as
\begin{eqnarray}
W^{tel}_{av} = \frac{2-p}{2}
\label{telfidwerner} 
\end{eqnarray}
from which it easily follows that in the limit of the maximally entangled
pure state ($p=1$) the channel is ideal, while for the maximally mixed
state ($p=0$), the fidelity falls below the classical limit of $2/3$.

In Fig.1 we plot the average teleportation fidelity $W^{tel}_{av}$
versus the channel parameter $p$.
A similar plot is also provided for the average teleportation fidelity
$F^{tel}_{av}$ corresponding to the non-orthogonal entangled channel
with respect to the parameter $r$ which also ranges between $0$ and $1$.
Note that in the region where both the fidelities exceed the classical
limit of $2/3$, the non-orthogonal channel always outperforms the
Werner state. Since, the latter is an example of a maximally entangled
mixed state, it is thus apparent that using a non-orthogonal channel
leads to a better efficiency of teleportation in average, compared to a
noisy channel. 

\begin{figure}[h]
{\rotatebox{0}{\resizebox{11.0cm}{8.0cm}{\includegraphics{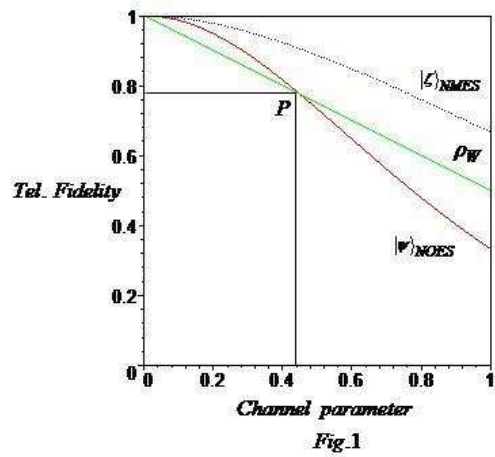}}}}
\caption{\footnotesize The teleportation fidelities of the three channels,
{\it viz.} the non-orthogonal state $|\psi\rangle_{NOES}$, the (mixed) Werner 
state $\rho^W$, and the 
non-maximally entangled pure state $|\xi\rangle_{NMES}$ are plotted versus
their respective channel parameters $r$, $p$ and $s$ for comparison. 
The teleportation efficiency
of $|\psi\rangle_{NOES}$ is better compared to  $\rho^W$ in the parameter range
where the classical limit of $2/3$ is exceeded. However, $|\xi\rangle_{NMES}$
outperforms the other two throughout.}
\end{figure}

Let us now consider an orthogonal non-maximally entangled state
as the channel for the teleportation of a single qubit state (\ref{input}).
This non-maximally entangled state can be written as
\begin{eqnarray}
|\zeta\rangle_{ab}=u|01\rangle+v|10\rangle\label{nonmax}
\end{eqnarray}
where $u^{2}+v^{2}=1$.
Here we note  the following facts: (i) if $u=0$ then the
two-qubit state (\ref{nonmax}) becomes separable, (ii) if
$u=\frac{1}{\sqrt{2}}$ then the two-qubit state reduces to a
maximally entangled state, and (iii) if $0<u<\frac{1}{\sqrt{2}}$
then the state is a non-maximally entangled state.
The parameter $u$
is a measure of non-maximality. Our purpose is to compare the efficiency
of this channel with the non-orthogonal channel studied in Section II. 
In order to
keep the two parameters $u$ (for the non-maximally entangled channel)
and $r$ (for the non-orthogonal channel) on the same footing, we have
to re-scale the parameter $u$ in such a way that it
can assume values between $0$ and $1$.
The re-scaled parameter $s$ can be written as
\begin{eqnarray}
s= 1-\sqrt{2}u\label{scaling}
\end{eqnarray}
Hence, $s=0$ and $s=1$ denote the maximally entangled state and
separable state, respectively. All other values of the parameter
$s$ lying between $0$ and $1$ correspond to  non-maximally entangled states.

\begin{figure}[h]
{\rotatebox{0}{\resizebox{11.0cm}{7.5cm}{\includegraphics{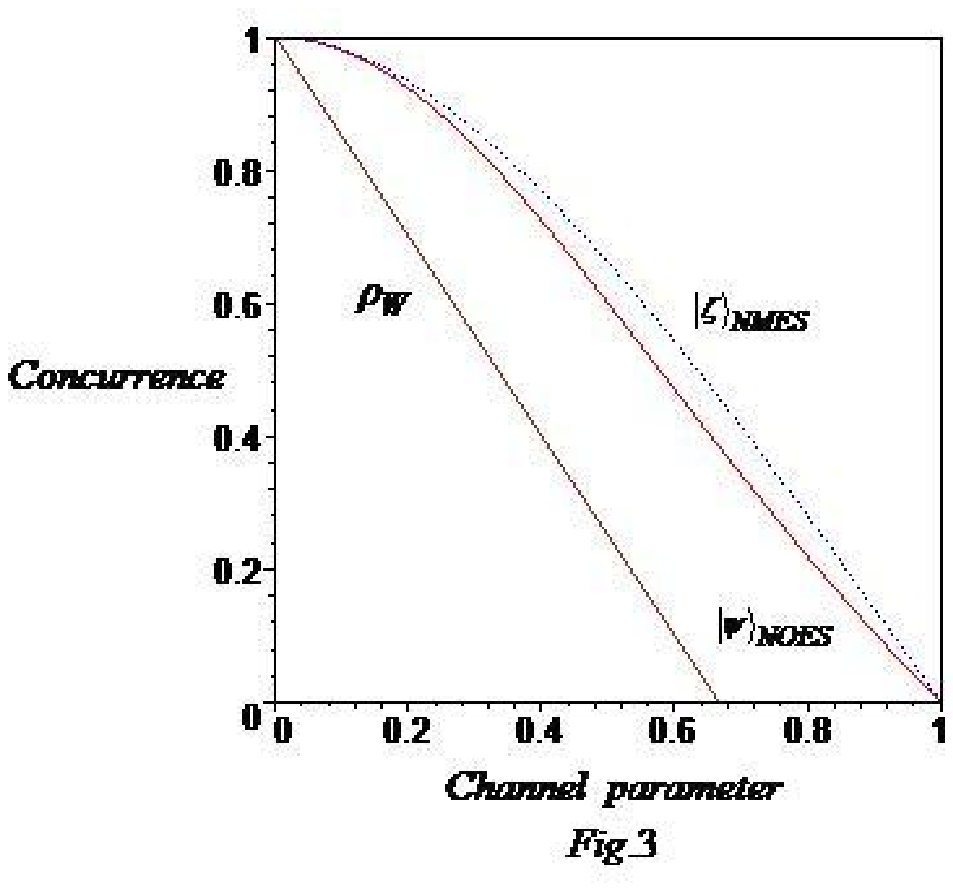}}}}
\caption{\footnotesize The magnitude of entanglement of the three channels,
{\it viz.} the non-orthogonal state $|\psi\rangle_{NOES}$, the (mixed) Werner 
state $\rho^W$, and the 
non-maximally entangled pure state $|\xi\rangle_{NMES}$ are plotted versus
their respective channel parameters $r$, $p$ and $s$ for comparison. The state 
$|\xi\rangle_{NMES}$ is more entangled compared to 
$|\psi\rangle_{NOES}$ which in turn is more entangled than $\rho^W$.}
\end{figure}

Now, repeating the conventional teleportation protocol for
orthogonal system using non-maximally entangled state
(\ref{nonmax}) as a teleportation channel, we can obtain the
teleportation fidelity in terms of the parameter $s$ as
\begin{eqnarray}
G^{tel}=
1-2y^{2}(1-y^{2})(1-(1-s)\sqrt{2-(1-s)^{2}})\label{telnonmax}
\end{eqnarray}
Clearly, the teleportation fidelity $G^{tel}$ is input state
dependent. Hence, for some input state it gives a better fidelity
compared to some other state. So, as in the previous case for the 
non-orthogonal channel, it would be better to
consider the average fidelity. The average
teleportation fidelity over all input state is given by
\begin{eqnarray}
G^{tel}_{av}=
\frac{2+(1-s)\sqrt{2-(1-s)^{2}}}{3}\label{telavnonmax}
\end{eqnarray}
Note in Fig1. that this channel performs better teleportation compared
to the non-orthogonal entangled state. Such a result can be understood
by observing the magnitude of entanglement of these channels as functions
of their respective parameters. The concurrence of all the three channels
are plotted in Fig.2 which shows that the non-maximmaly entangled pure
state is more entangled for a given parameter value compared to both the
non-orthogonal as well as the mixed state. To complete thet argument as
to why the non-maximally entangled state is more efficient than the 
non-orthogonal state as a teleportation channel, in Fig.3 we plot
the average teleportation fidelity corresponding to $|\psi\rangle_{NOES}$
and $|\xi\rangle_{NMES}$  as  function of
their respective concurrences. It is seen that for a given magnitude of
entanglement the state $|\xi\rangle_{NMES}$ outperforms the state
$|\psi\rangle_{NOES}$ as a teleportation channel.

\begin{figure}[h]
{\rotatebox{0}{\resizebox{11.0cm}{7.5cm}{\includegraphics{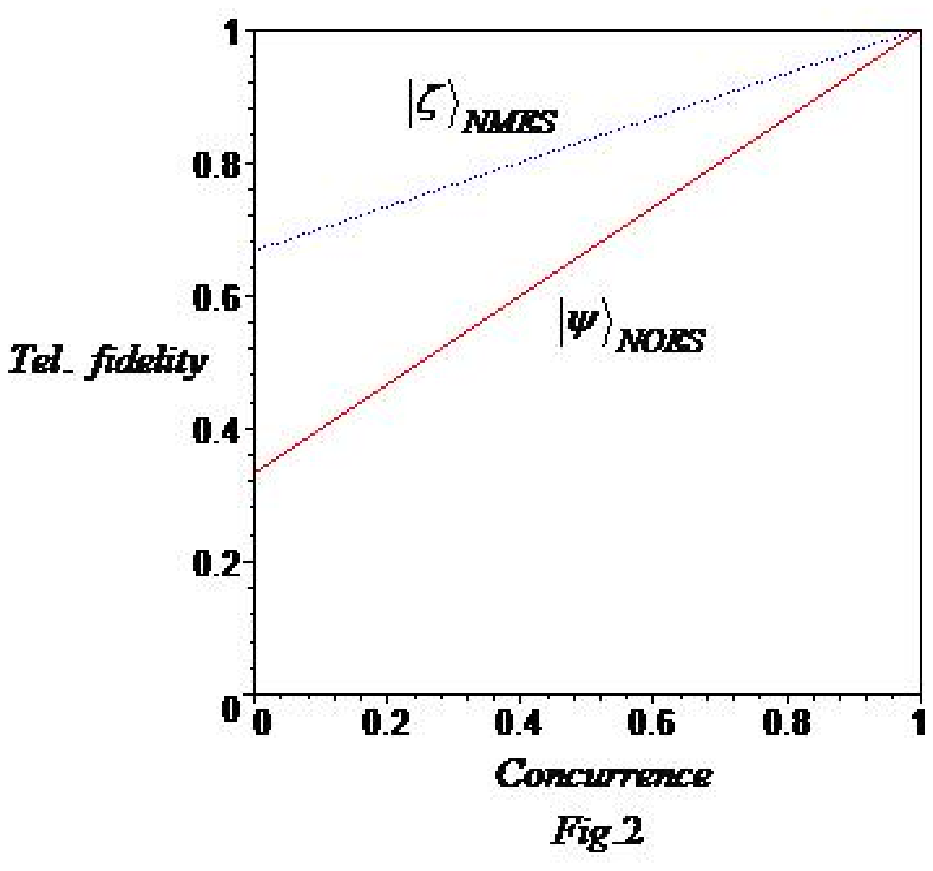}}}}
\caption{\footnotesize The average fidelity of teleportation of the 
the non-orthogonal channel $|\psi\rangle_{NOES}$, and the 
non-maximally entangled channel $|\xi\rangle_{NMES}$ are plotted versus
the amount of entanglement in these states.}
\end{figure}

\section{Efficiency of teleportation by mixed orthogonal and non-orthogonal states}

In the previous sections we have investigated a pure non-orthogonal entangled
state as a teleportation channel and compared its performance with that of
a pure non-maximally entangled state as well as with a 
maximally entangled mixed state. A natural question 
that might arise then is as
to how does a mixed non-orthogonal entangled channel fare in such a scheme
of teleportation. To address this issue, 
let us now consider a mixed state involving
non-orthogonal basis vectors, which can be defined as
\begin{eqnarray}
\hskip -0.2cm &&\rho^{N}=g |\psi\rangle\langle\psi|+\frac{1-g}{4}I, ~~~0\leq g
\leq1 {}\nonumber\\&&= \hskip -0.1cm
\left(\begin{matrix}{\frac{1-g}{4}+4N_{1}^{2}r^{2}g &
\frac{2gN_{1}^{2}\langle\alpha|\beta\rangle}{N_{G}} &
\frac{2gN_{1}^{2}\langle\alpha|\beta\rangle}{N_{G}} & 0\cr
\frac{2gN_{1}^{2}\langle\alpha|\beta\rangle^{*}}{N_{G}} &
\frac{1-g}{4}+g(\frac{N_{1}}{N_{G}})^{2}&g(\frac{N_{1}}{N_{G}})^{2}&0
\cr \frac{2gN_{1}^{2}\langle\alpha|\beta\rangle^{*}}{N_{G}} &
g(\frac{N_{1}}{N_{G}})^{2}
&\frac{1-g}{4}+g(\frac{N_{1}}{N_{G}})^{2}&0 \cr 0 &0 & 0 &
\frac{1-g}{4}}\end{matrix}\right)\label{non-ortho werner}
\end{eqnarray}
where
$|\psi\rangle=\frac{N_{1}}{N_{G}}(|0\rangle|1\rangle+|1\rangle|0\rangle)
+2N_{1}\langle\alpha|\beta\rangle|0\rangle|0\rangle$,
$N_{1}=\frac{1}{\sqrt{2(1+r^{2})}}$,
$N_{G}=\frac{1}{\sqrt{1-r^{2}}}$,
$\langle\alpha|\beta\rangle=re^{i\theta}$ and $I$ is the identity
operator.

For two-qubit systems, the negativity  defined as twice the largest
negative eigenvalue of the partially transposed density matrix, may 
be used to quantify
entanglement. The
negativity of $\rho^{N}$ is given by
\begin{eqnarray}
E_{N}=2max(0,-\lambda_{N}) \label{def.negativity}
\end{eqnarray}
where $\lambda_{N}$ denotes the negative eigenvalue of $(\rho^{N})^{PT}$.
Here we have
\begin{eqnarray}
E_{N}= \frac{g(3-r^2)-(1+r^2)}{2(1+r^2)},~~~\textrm{when}~~
\frac{1+r^{2}}{3-r^{2}}<g\leq1 \label{negativity}
\end{eqnarray}
For an entangled state $\rho^{N}$, the parameter $g$ can be written
as
\begin{eqnarray}
g=\frac{1+r^{2}}{3-r^{2}}+\epsilon,~~\epsilon>0 \label{parameter}
\end{eqnarray}
If $\epsilon<0$, the state $\rho^{N}$ is separable. Note that the condition
$0 \le g \le 1$ does not hold good when
$\epsilon>\frac{2}{3}$.

Now, it is known \cite{horodecki} 
that any mixed $spin-\frac{1}{2}$ state $\rho$ is
useful for (standard) teleportation if and only if
\begin{eqnarray}
\nu(\rho)=\sum_{i=1}^{3}\sqrt{u_{i}}>1 \label{tel.cond.}
\end{eqnarray}
For the two-qubit state $\rho^{N}$, $\nu(\rho^{N})$ is given by
\begin{eqnarray}
\nu(\rho^{N})=1+\frac{(3-r^{2})\epsilon}{1+r^{2}} \label{non-ortho
tel.cond.}
\end{eqnarray}
It follows that the two-qubit state described by the density 
matrix $\rho^{N}$ can
always be used as a teleportation channel when
$0<\epsilon<\frac{2}{3}$. The efficiency of the teleportation
channel is measured by the average teleportation fidelity which is given
by
\begin{eqnarray}
f_{av}^{tel}(\rho^{N})&&=\frac{1}{2}[1+\frac{\nu(\rho^{N})}{3}]
{}\nonumber\\&&=\frac{2}{3}+\frac{(3-r^{2})\epsilon}{6(1+r^{2})}
\label{tel.fid.}
\end{eqnarray}
For a given value of $\epsilon$ (or $g$), from the above expression one may 
compute the fidelity in terms of the non-orthogonality parameter $r$.

\begin{figure}[h]
{\rotatebox{0}{\resizebox{11.0cm}{7.5cm}{\includegraphics{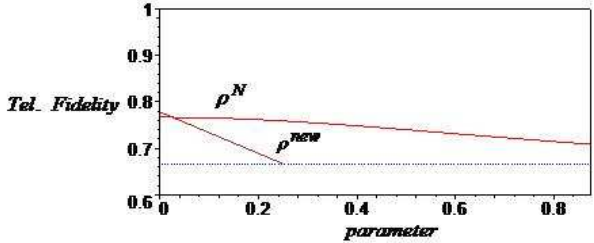}}}}
\caption{\footnotesize The average fidelity of teleportation for the 
the mixed non-orthogonal channel $f_{av}^{tel}(\rho^{N})$ 
(with $\epsilon = 0.2$),  and the 
non-maximally entangled mixed channel $f^{tel}_{av}(\rho_{new})$ 
are plotted versus
their respective channel parameters $r$ and $p$. The horizontal dotted 
line corresponds 
to the classical teleportation fidelity of $2/3$.}
\end{figure}

Next, we would like to compare the performance as teleportation channel
of the above mixed non-orthogonal entangled state with a mixed non-maximally
entangled state. As an example of the latter, we choose a state proposed
recently \cite{qic} given by a convex combination of a separable density matrix
$\rho^{G}_{12}=Tr_{3}(|GHZ\rangle_{123})$ and an inseparable
density matrix $\rho^{W}_{12}=Tr_{3}(|W\rangle_{123})$ as
\begin{eqnarray}
\rho_{new}=p\rho^{G}_{12}+(1-p)\rho^{W}_{12},~~~~0\leq p \leq 1
\label{T1}
\end{eqnarray}
where
$|GHZ\rangle$ and $|W\rangle$ denote the three-qubit GHZ-state\cite{ghz}
and the W-state\cite{wstate} respectively. This construction is somewhat
similar in spirit to the Werner state which is a convex combination of
a maximally mixed state and a maximally entangled pure state. Note that
the GHZ state and the W state are two qubit separable and
inseparable states, respectively, when a qubit is lost from the corresponding
three qubit states. The state $\rho_{new}$ was studied as a teleportation
channel in \cite{qic} where it was observed that it leads to a better
teleportation efficiency compared to some other non-maximally entangled 
mixed states. The average teleportation fidelity corresponding to $\rho_{new}$
is given by
\begin{eqnarray}
f^{tel}_{av}(\rho_{new})= \frac{7-4p}{9},~~~~0\leq p <\frac{1}{4}
 \label{T6}
\end{eqnarray}
with $\frac{2}{3}<f^{tel}_{av}(\rho_{new})\leq \frac{7}{9}$. The state
$\rho_{new}$ cannot be used as a efficient teleportation channel for 
$p \ge 1/4$, since in this case the teleportation fidelity falls below the 
classical fidelity.

\begin{figure}[h]
{\rotatebox{0}{\resizebox{11.0cm}{7.5cm}{\includegraphics{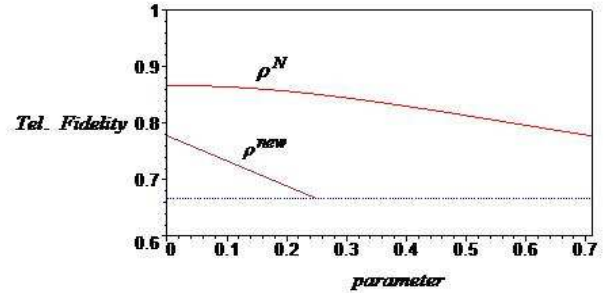}}}}
\caption{\footnotesize The average fidelity of teleportation for the 
the mixed non-orthogonal channel $f_{av}^{tel}(\rho^{N})$ 
(with $\epsilon = 0.4$),  and the 
non-maximally entangled mixed channel $f^{tel}_{av}(\rho_{new})$
are plotted versus
their respective channel parameters $r$ and $p$. The horizontal dotted 
line corresponds 
to the classical teleportation fidelity of $2/3$.}
\end{figure}

Finally, we present the comparison of the average teleportation fidelities
between the non-orthogonal mixed channel $\rho^N$ and the non-maximally
entangled mixed channel $\rho_{new}$. Their fidelities are plotted 
 versus the respective channel parameters $r$ and $p$
 for two different
values of the parameter $\epsilon$ in Fig.4 and Fig.5.
Note that the non-orthogonal mixed channel could lead to a better
output compared to the non-maximally entangled mixed channel for a range
of parameter values. This feature is different from the case of pure
states studied in the earlier section where we saw that the non-maximally
entangled pure channel is always more efficient compared to the 
non-orthogonal pure channel.

\section{Conclusions}

The motivation for this work has been to investigate the utility of
non-orthogonal entangled states as resource for performing the teleportation
of an unknown qubit. To this end we have first extended the standard
teleportation protocol \cite{bbcjpw} to the case of non-orthogonal
basis states. Since the non-orthogonal channel is less entangled than
the corresponding orthogonal case, teleportation through it leads to
a loss of fidelity. We have obtained the expression for the 
average teleportation fidelity corresponding to a non-orthogonal channel
that is idependent of the input states.
Subsequently, we have performed a comparison of the teleportation efficiency
of the non-orthogonal channel with other non-ideal channels corresponding
to mixed as well as non-maximally entangled states. We have also presented
a relative study of the teleportation fidelity of mixed states.

Our analysis leads to several interesting results. We first observe that
a non-orthogonal entangled channel could be more efficient as a teleportation
resource compared to a noisy channel. For the latter we consider the example 
of the Werner channel which is a maximally entangled mixed state. This result
follows from the fact that for a given range of channel parameter values, the 
Werner state,
in spite of being a maximally entangled mixed state, is still less entangled
than the corresponding non-orthogonal state. Our next comparative study
pertaining to a non-maximally entangled pure state shows that the latter
is always more efficient for performing teleportation compared to the 
non-orthogonal entangled state. We have finally presented a comparison
of mixed states in the orthogonal and non-orthogonal basis. Here we
compare the average teleportation fidelity of a non-orthogonal mixed
state with that of a non-maximally entangled mixed state \cite{qic}. 
Contrary to the
case of pure states, we find here that a suitable choice of parameter
values could lead to the non-orthogonal channel performing better
teleportation than the non-maximally entangled mixed state.  It would
be interesting to extend our study to the case of higher dimensions and
multipartite states. Such investigations could be useful for devising
channels for practical teleportation where it is almost impossible
to work with  ideal channels.

{\it Acknowledgements:} ASM, DH and PJ acknowledge support from the DST 
Project
SR/S2/PU-16/2007. DH thanks the Centre for Science and Consciousness, Kolkata,
for support. PJ acknowledges support from the INSA Summer Project 
program 
and thanks SNBNCBS for the hospitality provided during his visit.

\end{document}